\documentclass[aps,prd,groupedaddress,showpacs,twocolumn]{revtex4}
\usepackage{graphicx}
\usepackage{dcolumn}
\usepackage{bm}
\begin{document}

\title{Hidden local symmetry and the reaction $e^+e^-\to\pi^+\pi^-\pi^+\pi^-$ at energies $\sqrt{s}\leq$ 1 GeV.}

\author{N.~N.~Achasov}
\email[]{achasov@math.nsc.ru}
\affiliation{Laboratory of
Theoretical Physics, S.~L.~Sobolev Institute for Mathematics,
630090, Novosibirsk, Russian Federation
}%
\author{A.~A.~Kozhevnikov}
\email[]{kozhev@math.nsc.ru} \affiliation{Laboratory of
Theoretical Physics, S.~L.~Sobolev Institute for Mathematics, and
Novosibirsk State University, 630090, Novosibirsk, Russian
Federation}

\date{\today}
\begin{abstract}
Based on  the generalized hidden local symmetry as the chiral
model of pseudoscalar, vector, and axial vector mesons, the
excitation curve of the reaction $e^+e^-\to\pi^+\pi^-\pi^+\pi^-$
is calculated for  energies in the interval  $0.65\leq
\sqrt{s}\leq1$ GeV. The theoretical predictions are compared to
available data of CMD-2 and BaBaR.  It is shown that the inclusion
of heavy isovector resonances $\rho(1450)$ and $\rho(1700)$ is
necessary for reconciling calculations with the data. It is found
that at $\sqrt{s}\approx1$ GeV the contributions of the above
resonances are much larger, by the factor of 30, than the
$\rho(770)$ one, and are amount to a considerable fraction
$\sim0.3-0.6$ of the latter at $\sqrt{s}\sim m_\rho$.
\end{abstract}
\pacs{11.30.Rd;12.39.Fe;13.30.Eg}

\maketitle

\section{Introduction.}
\label{intro}~

The theory aimed at describing low energy hadron processes should
be formulated in terms of effective colorless degrees of freedom
\cite{weinberg79}. They are introduced on the basis of
spontaneously broken chiral symmetry $SU(3)_L\times SU(3)_R$ which
is the symmetry of QCD Lagrangian relative independent global
rotations of right and left fields of approximately massless
$u,d,s$ quarks. The pattern of the spontaneous breaking of the
above approximate symmetry is $SU(3)_L\times SU(3)_R\to
SU(3)_{L+R}$, where $SU(3)_{L+R}$ is the well known flavor $SU(3)$
symmetry. According to  Goldstone theorem, spontaneous breaking of
global symmetry results in appearance of massless fields. In the
present case,  they are light $J^P=0^-$ $\pi^+$, $\pi^-$, $\pi^0$,
$K^+$, $K^0$, $K^-$, $\bar K^0$, $\eta$. The transformation law of
these fields fixes the Lagrangian of interacting Goldstone mesons:
\begin{equation}
{\cal L}_{\rm GB}=\frac{f^2_\pi}{4}\mbox{Tr}\left(\partial_\mu
U\partial_\mu U^\dagger\right)+\cdots,\label{lgb}\end{equation}
where
\begin{equation}
U=\exp\left(i\Phi\sqrt{2}/f_\pi\right),\label{u}\end{equation}with
$$\Phi=\left(
\begin{array}{ccc}
\frac{\pi^0}{\sqrt{2}}+\frac{\eta_8}{\sqrt{6}}&\pi^+&K^+\\
\pi^-&-\frac{\pi^0}{\sqrt{2}}+\frac{\eta_8}{\sqrt{6}}&K^0\\
K^-&\bar K^0&-\frac{2\eta_8}{\sqrt{6}}
\end{array}\right),\;$$is the matrix of pseudoscalar meson octet,
and $f_\pi=92.4$ MeV is the pion decay constant. Upon adding the
term $\propto m^2_0\mbox{Tr}(U+U^\dagger)$ which explicitly breaks
chiral symmetry, the  Goldstone bosons become massive.

Pseudoscalar mesons are produced via vector resonances, so the
problem appears as how should one  include vector mesons in a
chiral invariant way? This problem was studied in a number of
papers, see, for example, Ref.~
\cite{weinberg68,ccwz,wz,schechter84,meissner88,ecker89,birse96}
and references therein. However, among  the  models aimed at the
description of interactions of the pseudoscalar mesons  with the
low lying vector and axial vector ones the most elegant  is the
generalized hidden local symmetry (GHLS) model
\cite{bando85,bando88a,bando88}. It relates all coupling constants
to only the pion decay constant $f_\pi$ and $g_{\rho\pi\pi}$, and
accounts for anomalous processes in a way that does not break low
energy theorems. Strikingly, but this very popular model was not
scrutinized in the processes with sufficiently soft pions where
one can rely on the tree approximation. The fact is that testing
chiral models of the vector meson interactions with Goldstone
bosons is really difficult problem because in the well studied
decays $\rho\to2\pi$, $\omega\to3\pi$ final  pions are not soft
enough to rely on lowest derivative tree effective Lagrangian.
Multiple pion decays are most promising because pions are soft.

The aim of the present paper is to confront the generalized hidden
local symmetry model \cite{bando85,bando88a,bando88} with
available data on the reaction $e^+e^-\to\pi^+\pi^-\pi^+\pi^-$
taken by CMD-2  \cite{cmd2} and  BaBar \cite{babar}
collaborations. The final state $\pi^+\pi^-\pi^+\pi^-$  can be
produced via intermediate $\rho(770)$ meson. In the framework of
chiral approach the $\rho(770)\to4\pi$ decay width was evaluated
in the papers
\cite{rittenberg69,bramon93,kuraev95,plant96,ach00a,ach00b,ach05}.
As was emphasized in Ref.~\cite{ach00a,ach00b},  because of a too
rapid growth with energy, the $\rho(770)\to4\pi$ decay  width
evaluated at $\sqrt{s}=m_\rho$ is not adequate characteristic of
the chiral dynamics, and one should study the excitation curve of
the process $\rho\to4\pi$ in such reactions as $e^+e^-$
annihilation, $\tau$ decays, photoproduction  etc. The
corresponding excitation curves were calculated  in
Ref.~\cite{ach00a,ach00b} in the chiral model which neglects the
$a_1(1260)$ contribution and under assumption of the resonant
mechanism $e^+e^-\to\rho(770)\to4\pi$ and similar in case of other
mentioned reactions.

The material is arranged as follows. Section \ref{ghls} is devoted
to the exposition of the low momentum expansion of the generalized
hidden local symmetry model lagrangian necessary for obtaining the
coupling constants of $\rho(770)$ meson and the virtual photon to
the  state $\pi^+\pi^-\pi^+\pi^-$. The amplitude of the reaction
$e^+e^-\to\pi^+\pi^-\pi^+\pi^-$ with the necessary lowest number
of derivatives is given in section \ref{amplitude}. Section
\ref{results} contains the results of the evaluation of the energy
dependence of the reaction $e^+e^-\to\pi^+\pi^-\pi^+\pi^-$ in GHLS
and the comparison of the calculations  with  the data
\cite{cmd2,babar} on the reaction $e^+e^-\to\pi^+\pi^-\pi^+\pi^-$.
The $\rho^\prime$, $\rho^{\prime\prime}$ contributions necessary
for reconciling the calculation with the data are studied in the
same section. Section \ref{discussion} is devoted to the
discussion of the results and to the comparison of GHLS approach
with different models exploited by other authors in order to
describe the reaction $e^+e^-\to\pi^+\pi^-\pi^+\pi^-$. Our
conclusions are stated in section \ref{conclusion}. The divergence
equation of the axial vector current allowing for external
electromagnetic field and its matrix element pertinent for the
reaction $e^+e^-\to\pi^+\pi^-\pi^+\pi^-$ in the hidden local
symmetry model is studied in Appendix. This is necessary for
studying the Adler limit $q_{\mu a}\to(0,0,0,0)$ ($q_{\mu a}$
being the four-momentum of any of the final pions) \cite{adler} of
the $e^+e^-\to\pi^+\pi^-\pi^+\pi^-$ reaction amplitude.

\section{Fixing coupling constants in GHLS.}\label{ghls}
~

The  virtue of generalized hidden local symmetry (GHLS) model
\cite{bando88,bando85,bando88a} is that, in the tree
approximation, there are no free parameters in the non-anomalous
sector except the masses of $\rho$ and $a_1$ mesons and the gauge
coupling constant $g=g_{\rho\pi\pi}$ determined from the
$\rho\to\pi^+\pi^-$ decay width, provided $f_\pi$ is known. All
couplings including $a_1\rho\pi$ and the direct $a_13\pi$, are
fixed by such natural requirements as vector meson dominance,
absence of higher derivative $\rho\pi\pi$ coupling, the
Kawarabayashi-Suzuki-Riazzuddin- Fayyazuddin (KSRF) relation
\cite{ksrf}
\begin{equation}
2g^2_{\rho\pi\pi}f^2_\pi=m^2_\rho,\label{KSRF}\end{equation} etc.
GHLS model was used in Ref.~\cite{ach05} devoted, in particular,
to the evaluation of the $\rho\to4\pi$ decay width at
$\sqrt{s}=m_\rho$. It is important that the electro-weak sector is
included into the framework independently of the strong
interacting one. This permits one to take into account contact
vertices  $\gamma^\ast\to4\pi$, $\gamma^\ast\to a_1\pi$ which
include the virtual photon $\gamma^\ast$, and the analogous ones
with the replacement $\gamma^\ast\to W^-$ in case of $\tau^-$
decays.

In order  to calculate the $e^+e^-\to\pi^+\pi^-\pi^+\pi^-$
excitation curve in the framework of GHLS model and to compare the
result with existing data of CMD-2 \cite{cmd2} and BaBaR
\cite{babar}, we use  recent calculations of the $\rho\to4\pi$
decay amplitudes \cite{ach05} and add them with the above
mentioned contact non-resonant terms whose explicit form is found
here.

In order to demonstrate the fixing of the coupling constants in
GHLS model, let us give the expressions for the  interaction
lagrangians following the parameters choice made in
Ref.~\cite{ach05} where the  necessary notations and  details can
be found. The boldface characters refer hereafter to the isotopic
vectors. The expressions include the following pieces.

(i) The simple hidden local symmetry (HLS) contribution arising in
case of neglecting the $a_1(1260)$ contribution
\begin{eqnarray}
{\cal L}_{\rm HLS}&=&\frac{m^2_\pi}{24f^2_\pi}{\bm\pi}^4+
\frac{1}{12f^2_\pi}[{\bm\pi}\times\partial_\mu{\bm\pi}]^2
+\nonumber\\&&g\left(1-\frac{{\bm\pi}^2}{12f^2_\pi}\right)
({\bm\rho}_\mu\cdot[{\bm\pi}\times\partial_\mu{\bm\pi}]).
\label{inthls}
\end{eqnarray}
It generates the $\pi\to3\pi$ vertices and the contact
$\rho\to4\pi$ one.

(ii) The term responsible for the decay
$a_1\to\rho\pi+3\pi\to3\pi$
\begin{eqnarray}
{\cal
L}_{a_1\rho\pi+a_13\pi}&=&-\frac{1}{f_\pi}(\partial_\mu{\bm\rho}_\nu
-\partial_\nu{\bm\rho}_\mu)\cdot[{\bm
a}_\mu\times\partial_\nu{\bm\pi}]-\nonumber\\&&\frac{1}{2f_\pi}(\partial_\mu{\bm
a}_\nu -\partial_\nu{\bm a}_\mu)\cdot[{\bm
\rho}_\mu\times\partial_\nu{\bm\pi}]-\nonumber\\&&\frac{1}{8gf^3_\pi}[{\bm
a}_\mu\times\partial_\nu{\bm\pi}]\cdot[\partial_\mu{\bm\pi}\times\partial_\nu{\bm\pi}]
-\nonumber\\&&\frac{1}{4gf^3_\pi}\partial_\mu{\bm
a}_\nu\cdot[{\bm\pi}\times[\partial_\mu{\bm\pi}\times\partial_\nu{\bm\pi}]].
\label{a1rhopi}\end{eqnarray} It is essential that both
$a_1\rho\pi$ and the contact $a_13\pi$ terms are necessary for
fulfilling the Adler condition \cite{adler} in the chiral limit
$m_\pi\to0$, that is the vanishing of the amplitude at the
vanishing four-momentum of any final pion. This is the point of
departure of the present consideration from that of
Ref.~\cite{ecker} where the contact terms are absent but the
$a_1\rho\pi$ interaction vertex contains additional derivative as
compared to Eq.~(\ref{a1rhopi}) and is characterized by three
arbitrary parameters. To be more precise, the first two lines of
Eq.~(\ref{a1rhopi}) and the $a_1\rho\pi$ lagrangian in the paper
\cite{ecker} can be shown to be equivalent, but only on the mass
shells of {\it both} $a_1(1260)$ and $\rho(770)$ mesons.  In our
case these two resonances are off mass shells, so that restricting
by the first two lines    in Eq.~(\ref{a1rhopi}) would result in
breaking of the Adler condition for the $a_1\to 3\pi$ decay
amplitude. Non-resonant $a_1\to 3\pi$ terms written down in
Eq.~(\ref{a1rhopi}) restore chiral symmetry and the Adler
condition. The $a_1\rho\pi$ coupling in the paper \cite{ecker}
results in $a_1\to 3\pi$ decay amplitude which obeys the Adler
condition just due to its higher derivative form.

(iii) There are also the $\rho\to\rho\pi\pi$ and the higher
derivative contact $\rho\to4\pi$ terms arising due to the
procedure of diagonalization of the axial vector-pseudoscalar
mixing added with the counter terms \cite{bando88,bando88a}. They
are \cite{ach05}
\begin{eqnarray}
{\cal
L}_{\rho\rho\pi\pi+\rho4\pi}&=&-\frac{1}{16f^2_\pi}\left([{\bm
\rho}_\mu\times\partial_\nu{\bm\pi}]-[{\bm
\rho}_\nu\times\partial_\mu{\bm\pi}]\right)^2-\nonumber\\&&\frac{1}{8gf^4_\pi}[{\bm
\rho}_\mu\times\partial_\nu{\bm\pi}]\cdot[{\bm\pi}\times[\partial_\mu{\bm\pi}
\times\partial_\nu{\bm\pi}]].
\label{rhorhopipi}\end{eqnarray}Again, the contact term is
necessary for fulfilling the Adler condition of the corresponding
contribution to the $\rho\to4\pi$ decay amplitude \cite{ach05}.

(iv) The terms due to the direct photon coupling (${\cal A}_\mu$,
${\bm a}_\mu$ stand for the photon four-vector potential and
$a_1(1260)$ meson field, respectively) are given by
\begin{eqnarray}
{\cal L}_{\rm photon}&=&-e{\cal A}_\mu\left(2gf^2_\pi\rho^0_\mu-
\frac{\pi^+\pi^-}{2f^2_\pi}[{\bm\pi}\times\partial_\mu{\bm\pi}]_3
-\right.\nonumber\\&&\left.2g\rho^0_\mu\pi^+\pi^-+2gf_\pi[{\bm\pi}\times{\bm
a}_\mu]_3\right).\label{photon}\end{eqnarray}Notice that here are
given only the terms necessary for the $\pi^+\pi^-\pi^+\pi^-$
final state, and the contributions of the second order in electric
charge $e$ are neglected. The first, second,  third, and  fourth
terms in Eq.~(\ref{photon}) describe, respectively, the
$\gamma^\ast\to\rho^0$ transition, the contact
$\gamma^\ast\to\pi^+\pi^-\pi^+\pi^-$,
$\gamma^\ast\to\rho^0\pi^+\pi^-$, and $\gamma^\ast\to
a^\pm_1\pi^\mp$ vertices. One should have in mind that the contact
$\gamma^\ast\to\pi^+\pi^-$ and
$\gamma^\ast\to\pi^+\pi^-\pi^+\pi^-$ vertices cannot be
simultaneously eliminated in HLS \cite{fn0}.  See Appendix for the
discussion of the details of the direct pointlike contribution in
the hidden local symmetry model.

\section{The amplitude of the reaction
$e^+e^-\to\pi^+\pi^-\pi^+\pi^-$.} \label{amplitude}~

Let us represent the energy dependence of the
$e^+e^-\to\pi^+\pi^-\pi^+\pi^-$ reaction cross section in the form
\begin{equation}
\sigma_{e^+e^-\to4\pi}(s)=\frac{12\pi m^3_\rho\Gamma_{\rho
e^+e^-}(m_\rho) \Gamma^{\rm
eff}_{\rho\to4\pi}(s)}{s^{3/2}|D_\rho(q)|^2},
\label{curve}\end{equation}where the  leptonic width of the vector
meson $V$ on the mass shell looks as
\begin{equation}
\Gamma_{Ve^+e^-}(m_V)=\frac{4\pi\alpha^2m_V}{3f^2_V},\label{leptwidth}
\end{equation} and  $s=q^2$ is the total energy
squared in the center-of-mass system. The function   $\Gamma^{\rm
eff}_{\rho\to4\pi}(s)$  in Eq.~(\ref{curve}) can be evaluated with
the help of Eq.~(5.1) in Ref.~\cite{ach05}. For the purposes of
the present work it should be evaluated with the effective
$\rho\to4\pi$ decay amplitude $M^{\rm eff}_{\rho\to4\pi}\equiv
M^{\rm
eff}_{\rho_q\to\pi^+_{q_1}\pi^+_{q_2}\pi^-_{q_3}\pi^-_{q_4}}$
which includes both the resonant contribution
$e^+e^-\to\gamma^\ast\to\rho\to\pi^+\pi^-\pi^+\pi^-$ due to
Eq.~(\ref{inthls}), (\ref{a1rhopi}), and (\ref{rhorhopipi}) side
by side with the  contact one
$e^+e^-\to\gamma^\ast\to\pi^+\pi^-\pi^+\pi^-$ due to the terms
Eq.~(\ref{photon}). In the lowest order in electromagnetic
coupling constant this amplitude  is given by the expression
\begin{equation}
M^{\rm eff}_{\rho\to4\pi}=
\frac{g_{\rho\pi\pi}}{f^2_\pi}\epsilon_\mu(A_1q_{1\mu}+A_2q_{2\mu}+A_3q_{3\mu}+A_4q_{4\mu}),
\label{meff}\end{equation}where $\epsilon_\mu$ stands for the
polarization four-vector of the virtual $\rho$ meson, and
$A_a\equiv A_a(q_1,q_2,q_3,q_4)$, $a=1,2,3,4$ are dimensionless
invariant functions. Then, say, $A_1\equiv A_1(q_1,q_2,q_3,q_4)$
is given by the expression
\begin{widetext}
\begin{eqnarray}
A_1&=&-1+(1+\widehat{P}_{34})B_1,\nonumber\\
B_1&=&\frac{2}{D_\pi(q-q_1)}\left[\frac{m^2_\rho}{D_{\rho23}}(q_4,q_2-q_3)-(q_2,q_3)\right]
-D_\rho(q)\left(\frac{1}{D_{\rho14}}-\frac{1}{2m^2_\rho}\right)-
\frac{(1-\widehat{P}_{23})}{4D_{a_1}(q-q_1)}\times\nonumber\\&&
\left\{\frac{1}{D_{\rho23}}\left[4(q_2,q_4)(2q-q_1,q_3)-(2q-q_1,q_2)(q_4,q-q_1)+(2q-q_1,q_4)(q_2,q_4)\right]
-\right.\nonumber\\&&\left.\frac{1}{2m^2_\rho}(q_2,2q-2q_1+q_4)(2q-q_1,q_3)-
\frac{4(q_2,q_4)(q_2+q_3)^2}{m^2_{a_1}}\left[q^2+D_\rho(q)\right]\left(\frac{1}{D_{\rho23}}-
\frac{1}{8m^2_\rho}\right)\right\}-\nonumber\\&&\frac{3(q,q_2)-m^2_\pi-D_\rho(q)}{4D_{a_1}(q-q_2)}
\left[\frac{(q_4,4q_3-q_2+q)}{D_{\rho13}}-\frac{(q_4,2q-2q_2+q_3)}{2m^2_\rho}\right]+\frac{(q_4,q_2-q_3)}{4D_{\rho23}}
+\frac{(q_4,q_2+q_3)}{4m^2_\rho}-\nonumber\\&&
\frac{3(q,q_4)-m^2_\pi-D_\rho(q)}{4D_{a_1}(q-q_3)}
\left[\frac{(q_2,4q_3-q_4+q)}{D_{\rho13}}-\frac{(q_1,q_2-q_3)}{D_{\rho23}}-\frac{(q_3,2q-2q_4+q_2)}{2m^2_\rho}\right]
-\frac{(q_2,q_3)}{2D_{\rho14}},\label{A1}\end{eqnarray}
\end{widetext}
where $\widehat{P}_{ab}$ is the operator interchanging the pion
momenta $q_a\leftrightarrow q_b$, and $D_{\rho ab}\equiv
D_\rho(q_a+q_b)$ is the inverse propagator of $\rho$ meson with
the invariant mass squared $(q_a+q_b)^2$. The inverse propagator
of $\rho$ meson with the four-momentum $q$ and the invariant mass
$\sqrt{q^2}$ is taken in the form
\begin{equation}
D_\rho(q)=m^2_\rho-q^2-i\sqrt{q^2}\Gamma_\rho(\sqrt{q^2}),\label{Drho}\end{equation}
see Eqs.~(3.3)$-$(3.5) of Ref.~\cite{ach05} for explicit
expression of $\Gamma_\rho$. Notice, that the terms $\propto
D_\rho(q)$ in Eq.~(\ref{A1}) refer to the contact terms generated
by Eq.~(\ref{photon}).  The remaining notations are as follows.
$(P,Q)$ stands for invariant scalar product of two four-vectors
$P$ and $Q$, $D_\pi(p)=m^2_\pi-p^2$ is the inverse propagator of
pion, $m_\pi$ and $m_\rho$ are the masses of charged pion and
$\rho(770)$ meson whose values are taken from Ref.~\cite{pdg}. The
invariant amplitudes $A_{2,3,4}$ are obtained from  $A_1$ in
accord with the relations
\begin{eqnarray}
A_2&\equiv &A_2(q_1,q_2,q_3,q_4)=A_1(q_2,q_1,q_3,q_4),\nonumber\\
A_3&\equiv &A_3(q_1,q_2,q_3,q_4)=-A_1(q_3,q_4,q_1,q_2),\nonumber\\
A_4&\equiv &A_4(q_1,q_2,q_3,q_4)=-A_1(q_4,q_3,q_1,q_2).
\end{eqnarray}
The details of evaluation of $\Gamma_{\rho\to4\pi}(s)$ from
$M_{\rho\to4\pi}$, including the form of the $a_1$ propagator
$D^{-1}_{a_1}$ with the energy dependent width, are given in
Ref.~\cite{ach05}. In the present paper  we use the approximate
expression  for the energy dependent width $\Gamma_{a_1}(m)$ which
interpolates the  curve in Ref.~\cite{ach05} in the range
$0.6<m<0.86$ GeV, $\sqrt{s}\leq1$ GeV.  The numerical input for
$\Gamma_{a_1}(m)$ valid in this range  can be  represented by the
expression $\Gamma_{a_1}(m)\approx10^{8.47931m-9.07101}$, where
$\Gamma_{a_1}$ and $m$ are expressed in the units of  GeV. Note
that the approximate exponential form is just the  numerical
artifact due to very strong energy dependence arising via combined
action of the threshold factor in the phase space and the powers
of the pion momenta in the matrix element respecting the demands
of the chiral symmetry. In fact, the running $\Gamma_{a_1}$ can be
neglected at $m<0.7$ GeV because $\Gamma_{a_1}(m=0.7{\rm
GeV})=8\times10^{-4}$ GeV, while in order to reach 10 percent in
$D_{a_1}(m)$  it should amount to 0.1 GeV at the above energy.

\begin{figure}
\includegraphics[width=85mm]{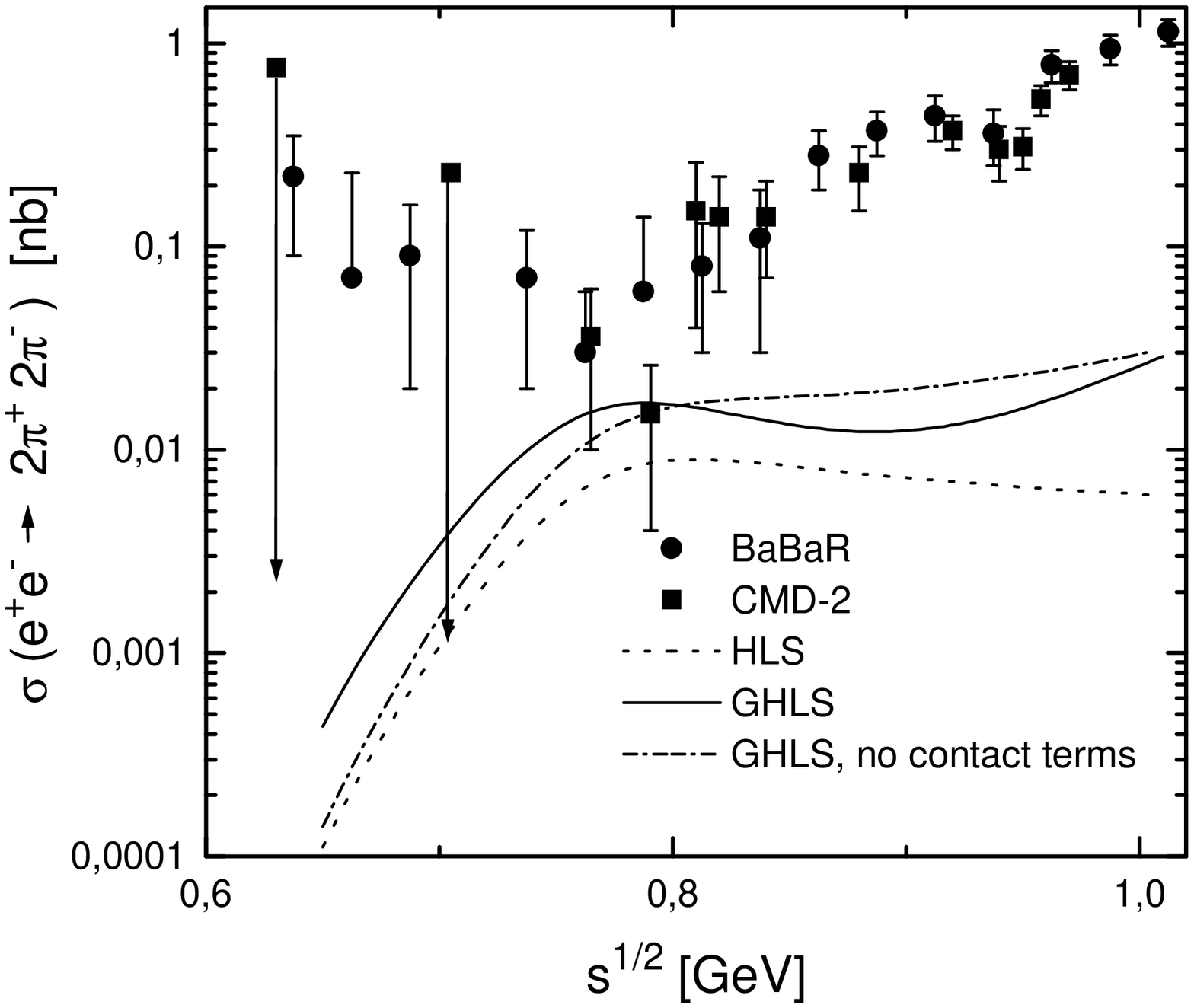}
\caption{\label{purehls}The energy dependence of the
$e^+e^-\to\pi^+\pi^-\pi^+\pi^-$ reaction cross section in the
generalized hidden local symmetry model, $m_{a_1}=1.23$ GeV. The
data are CMD-2 \cite{cmd2} and BaBaR \cite{babar}. "HLS" refers to
the lagrangian Eq.~(\ref{inthls}). "GHLS" refers to the model
based on lagrangian Eq.~(\ref{inthls}), (\ref{a1rhopi}),
(\ref{rhorhopipi}), and (\ref{photon}). "GHLS, no contact terms"
refers to the model without contact terms Eq.~(\ref{photon}).}
\end{figure}

The resonant contribution
$\gamma^\ast\to\rho\to\pi^+\pi^-\pi^+\pi^-$ in Eq.~(\ref{meff})
respects the requirement of chiral symmetry in that it vanishes at
the vanishing momentum $q_{a\mu}\to 0$ ($a=1,2,3,4$) of any final
pion \cite{adler}. However, the terms due to the direct
$\gamma^\ast\to\pi^+\pi^-\pi^+\pi^-$ contribution in
Eq.~(\ref{meff}) do not vanish in the above limiting cases. This
is the consequence of the breaking of the chiral symmetry by
electromagnetic field. As is shown in Appendix, the terms in the
amplitude Eq.~(\ref{meff}) surviving in the limit $q_{a\mu}\to 0$,
correspond to the matrix elements of the divergence of the axial
current.

\section{Comparison with the CMD-2 and BaBaR data.}
\label{results} ~

The results of evaluation of the $e^+e^-\to\pi^+\pi^-\pi^+\pi^-$
reaction cross section in the model specified by the lagrangians
Eq.~(\ref{inthls}), (\ref{a1rhopi}), (\ref{rhorhopipi}), and
(\ref{photon}) are shown in Fig.~\ref{purehls}. The curves are
obtained in the case $m_{a_1}=1.23$ GeV; the results for the mass
$m_{a_1}=1.09$ GeV look qualitatively the same. One can see that
the model is unable to reproduce the magnitude of the cross
section at energies $\sqrt{s}>0.8$ GeV.

Let us include the contributions of heavier resonances
$\rho^\prime\equiv\rho(1450)$ and
$\rho^{\prime\prime}\equiv\rho(1700)$ trying to explain the cross
section magnitude at $\sqrt{s}\geq0.8$ GeV. Since the far left
shoulder of the $\rho^\prime,\rho^{\prime\prime}$ resonance peaks
is not a proper place to their study, we choose the simplest
parametrization consisting of the Breit-Wigner resonance shape
with the constant widths and masses $m_{\rho^\prime}=1.459$ GeV,
$\Gamma_{\rho^\prime}=0.147$ GeV, $m_{\rho^{\prime\prime}}=1.72$
GeV, $\Gamma_{\rho^{\prime\prime}}=0.25$ GeV taken from
Ref.~\cite{pdg} and neglect the $\rho(770)-\rho(1450)-\rho(1700)$
mixing due to their common decay modes \cite{ach97,fn1}. In
addition, we take into account the model of $a_1\pi$ dominance  in
the $\rho^\prime,\rho^{\prime\prime}\to4\pi$ decay dynamics
proposed in Ref.~\cite{cmd2_99}, but modify it in order to make
the corresponding terms to obey the Adler condition, see
Eq.~(\ref{a1rhopi}) and Ref.~\cite{ach05}. Then taking into
account the $\rho^\prime,\rho^{\prime\prime}$ resonance
contributions results in the factor
\begin{equation}
R(s)=\left|1+\frac{D_\rho(q)}{1+r(s)}\left[\frac{x_{\rho^\prime}}{D_{\rho^\prime}(q)}+
\frac{x_{\rho^{\prime\prime}}}{D_{\rho^{\prime\prime}}(q)}\right]\right|^2,
\label{rhoprime}
\end{equation}multiplying the right hand side of
Eq.~(\ref{curve}), where $D_V(q)=m^2_V-s-im_V\Gamma_V$,
$V=\rho^\prime,\rho^{\prime\prime}$, $s=q^2$; $x_{\rho^\prime}$
and $x_{\rho^{\prime\prime}}$ are free parameters to be determined
from fitting the data. The meaning of $x_{\rho^\prime}$ is that
\begin{equation}
x_{\rho^\prime}=\frac{g_{\gamma\rho^\prime}}{g_{\gamma\rho}}\frac{g
_{\rho^\prime\to a_1\pi\to4\pi}}{g_{\rho\to
a_1\pi\to4\pi}},\label{xpr1}\end{equation} analogously for
$x_{\rho^{\prime\prime}}$, where $g _{\rho^\prime\to
a_1\pi\to4\pi}$ etc. means the amplitude corresponding to the
specific intermediate state $a_1\pi$ followed by  both the
resonant $a_1\to\rho\pi\to3\pi$ and the direct transition
$a_1\to3\pi$ of the intermediate $a_1$ meson. Since $\rho$ and
$\rho^\prime$ are assumed here to have the similar coupling to the
state $a_1\pi$, the ratio Eq.~(\ref{xpr1}) is constant. As usual,
$g_{\gamma V}=em^2_V/f_V$ stand for the amplitude of the
photon-vector meson $V$ transition, and $f_V$ is related with the
leptonic width Eq.~(\ref{leptwidth}). The complex function $r(s)$
in Eq.~(\ref{rhoprime}) is the ratio of the amplitude with the
intermediate $a_1$ meson to one with no $a_1$ contribution. It
approximately takes into account the $a_1\pi$ dominance in the
four pion decay of heavy isovector resonances. We precalculate it
for the fitting purposes for the CMD-2 \cite{cmd2} and BaBaR
\cite{babar} data points $\sqrt{s}\leq1$ GeV in accord with the
relations
\begin{eqnarray}
r(s)&=&\left[\frac{\Gamma^{{\rm eff, no}a_1}_{\rho\to
4\pi}}{\Gamma_{\rho\to
a_1\pi\to4\pi}}\right]^{1/2}\exp(i\chi),\nonumber\\
\chi&=&\cos^{-1}\frac{\Gamma^{\rm eff}_{\rho\to4\pi}-\Gamma^{{\rm
eff, no}a_1}_{\rho\to 4\pi}-\Gamma_{\rho\to
a_1\pi\to4\pi}}{2\sqrt{\Gamma_{\rho\to a_1\pi\to4\pi}\Gamma^{{\rm
eff, no}a_1}_{\rho\to 4\pi}}},\label{r}\end{eqnarray} where
$\Gamma_{\rho\to a_1\pi\to4\pi}\equiv\Gamma_{\rho\to
a_1\pi\to4\pi}(s)$ is the $\rho^0\to\pi^+\pi^-\pi^+\pi^-$ decay
width due to the intermediate $a_1\pi$ state only, while
$\Gamma^{{\rm eff, no}a_1}_{\rho\to 4\pi}\equiv\Gamma^{{\rm eff,
no}a_1}_{\rho\to 4\pi}(s)$ is the effective width of the same
decay including all the contribution mentioned above except the
$a_1\pi$ one. Hence, the approximation of Eq.~(\ref{r})
corresponds to the  averaging over four pion phase space. The
approximation is necessary, because the direct evaluation would
take unacceptable long time for numerical calculations in the
fitting procedure.

The CMD-2 \cite{cmd2} and BaBaR \cite{babar} data are taken at
different apparatus, with different methods. The systematic
uncertainties are usually estimated rather subjectively and are
naturally different on each detector. So it is more correct to
treat different data sets separately. Although, at first sight,
two data sets seem to be compatible,  fitting them in the
framework of the single model  gives different central values of
the fitted parameters and $\chi^2/{\rm n.d.f.}$, see below.

The results of fitting the CMD-2 data are given in Table
\ref{rescmd2}.
\begin{table}
\caption{\label{rescmd2}The results of fitting CMD-2 data
\cite{cmd2}. }
\begin{ruledtabular}
\begin{tabular}{ccccc}
variant&$x_{\rho^\prime}$
&$x_{\rho^{\prime\prime}}$&$\chi^2/N_{\rm
d.o.f}$&$m_{a_1}$ [GeV]\\
\hline1& $-27.5\pm1.5$&$\equiv0$&15.4/10&1.23\\
      2& $\equiv0$&$-46.2\pm2.5$&15.4/10&1.23\\
      3& $96.8\pm1.5$&$-208.7\pm2.5$&14.5/9&1.23\\
      4& $-17.8\pm1.0$&$\equiv0$&15.7/10&1.09\\
      5& $\equiv0$&$-30.1\pm1.5$&15.4/10&1.09\\
      6& $72.5\pm1.0$&$-151.9\pm1.6$&14.7/9&1.09
\end{tabular}
\end{ruledtabular}
\end{table}
\begin{figure}
\includegraphics[width=85mm]{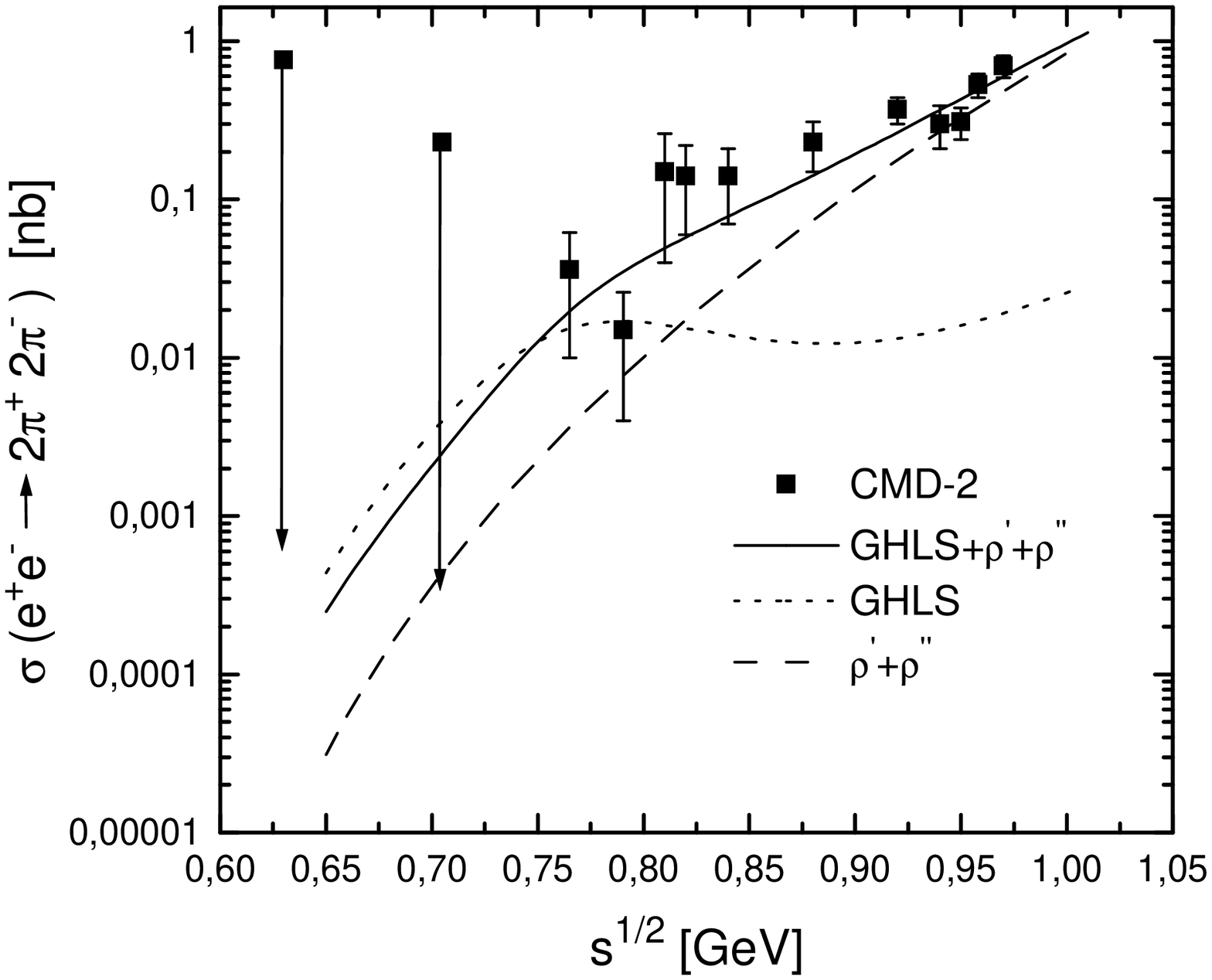}
\caption{\label{fitcmd2}The results of fitting the CMD-2 data
\cite{cmd2}.  "GHLS" refers to the model based on lagrangian
Eq.~(\ref{inthls}), (\ref{a1rhopi}), (\ref{rhorhopipi}), and
(\ref{photon}). See text for details.}
\end{figure}
The curves corresponding to the fit variant  3 are shown in
Fig.~\ref{fitcmd2}. This is the variant with two additional heavy
resonances $\rho^\prime$ and $\rho^{\prime\prime}$, and it is
indistinguishable from the variants with the single resonance
$\rho^\prime$ (variant 1) or $\rho^{\prime\prime}$ (variant 2),
both resulting in the same curves as the dashed one shown in
Fig.~\ref{fitcmd2}. However, variant 3 is based on the
destructively interfering large contributions of $\rho^\prime$ and
$\rho^{\prime\prime}$, so that each of the above (not shown) is
large as compared to their sum. Variants $4-6$ correspond to the
fits with the mass of $a_1$ meson $m_{a_1}=m_\rho\sqrt{2}=1.09$
GeV as given by Weinberg's relation and result in the same
corresponding curves not shown here. One can see that all the
fitting variants are not quite good. Nevertheless, we quote the
contribution of the sum $\rho^\prime+\rho^{\prime\prime}$ ( in
variant 3) or $\rho^\prime$ (variant 1) and $\rho^{\prime\prime}$
(variant 2) relative to the case of pure GHLS contribution (dotted
line in Fig.~\ref{fitcmd2}) to be 0.3 at $\sqrt{s}\approx m_\rho$
and 32 at $\sqrt{s}=1$ GeV. These numbers refer to the case
$m_{a_1}=1.23$ GeV. The case $m_{a_1}=1.09$ GeV results in almost
the same figures for above ratios.

The results of the similar analysis of the BaBaR data \cite{babar}
are presented in Table \ref{resbabar}.
\begin{table}
\caption{\label{resbabar}The results of fitting BaBaR data
\cite{babar}. }
\begin{ruledtabular}
\begin{tabular}{ccccc}
variant&$x_{\rho^\prime}$
&$x_{\rho^{\prime\prime}}$&$\chi^2/N_{\rm
d.o.f}$&$m_{a_1}$ [GeV]\\
\hline1& $-25.2\pm0.9$&$\equiv0$&32.6/16&1.23\\
      2& $\equiv0$&$-44.0\pm2.1$&29.3/16&1.23\\
      3& $273.2\pm1.4$&$-514.5\pm2.3$&11.2/15&1.23\\
      4& $-15.8\pm0.8$&$\equiv0$&35.0/16&1.09\\
      5& $\equiv0$&$-27.7\pm1.3$&31.8/16&1.09\\
      6& $198.5\pm1.0$&$-370.1\pm1.5$&11.2/15&1.09
\end{tabular}
\end{ruledtabular}
\end{table}
\begin{figure}
\includegraphics[width=85mm]{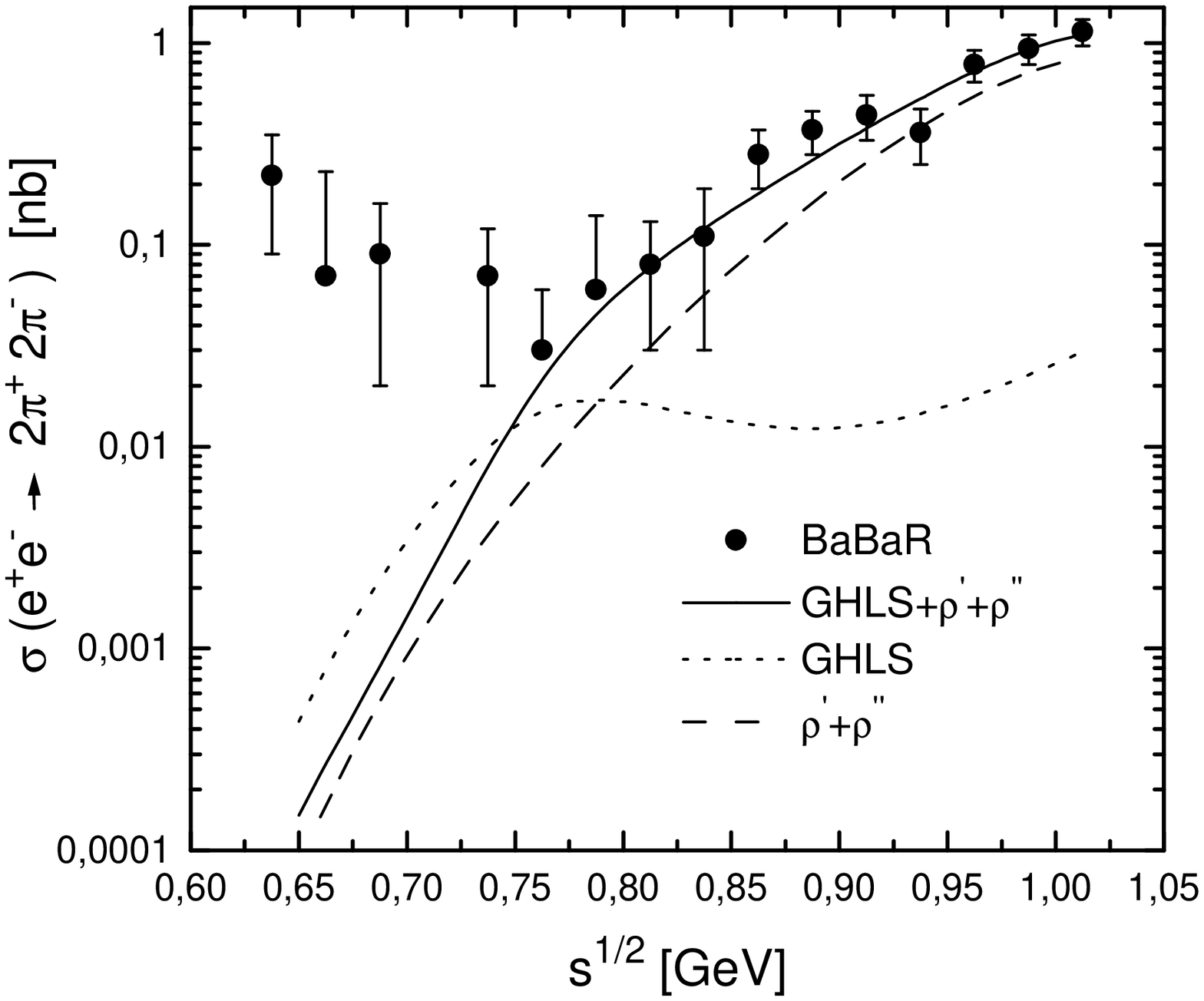}
\caption{\label{fitbabar}The same as in Fig.~\ref{fitcmd2}, but
for the BaBaR data \cite{babar}.}
\end{figure}
Contrary to the previous case, here the variants with the single
additional heavy resonance give a bad description. The fit chooses
two destructively interfering $\rho^\prime$ and
$\rho^{\prime\prime}$ resonances each coupled to $a_1\pi$ much
strongly than in the variants of the single heavy resonance. The
curves shown in Fig.~\ref{fitbabar} refer to variant 3 in
Table~\ref{resbabar} with $m_{a_1}=1.23$ GeV. The  contribution of
the sum $\rho^\prime+\rho^{\prime\prime}$ ( in variant 3) or
$\rho^\prime$ (variant 1) and $\rho^{\prime\prime}$ (variant 2)
relative to the case of pure GHLS contribution (dotted line in
Fig.~\ref{fitbabar}) to be 0.6 at $\sqrt{s}\approx m_\rho$ and 30
at $\sqrt{s}=1$ GeV. As in the case of the CMD-2 data, here the
variant 6 with $m_{a_1}=1.09$ GeV results in practically the same
corresponding curves and ratios.

\section{Discussion.}
\label{discussion}~

As is found  in the present paper, the $\rho^\prime$,
$\rho^{\prime\prime}$ contributions are large even at
$\sqrt{s}\simeq1$ GeV. So, our conclusions differ from the results
presented in Ref.\cite{cmd2,ecker,lichard}. Indeed, the
contribution of heavy resonances in Ref.~\cite{cmd2} is rather
small: the ratio of the $\rho^\prime$ to $\rho$ contributions is
0.02 at $\sqrt{s}=0.8$ GeV and grows to the figure of 0.15 at
$\sqrt{s}=1$ GeV \cite{cmd2}. Ref.~\cite{ecker,lichard} also point
to the possibility of describing the low energy data without the
$\rho^\prime$ contribution \cite{ecker} or with the small one
\cite{lichard}. We attribute this disagreement to the difference
among the models used in the present  analysis and in that of
Ref.~\cite{cmd2,cmd2_99,ecker,lichard}. Indeed, the
parametrization used by CMD-2 \cite{cmd2,cmd2_99} is based on
effective vertices provided by the isobar model, where    one
introduces all possible effective terms  allowed by Lorentz
invariance and G-parity. The amplitude  does not obey the demands
of chiral symmetry expressed in the property of the divergence of
the axial vector current. In contrast,  our parametrization is
much more restrictive since satisfies    requirements of chiral
symmetry. Hence, the strong chiral  cancellations  among different
terms in the    amplitude take place. This results in much
stronger $\rho^\prime$, $\rho^{\prime\prime}$ contributions. The
model in Ref.~\cite{lichard} is also  of the kind of effective
isobar model. It is purely phenomenological chiral-non-invariant
model in that part which concerns $a_1\rho\pi$ coupling. The
contact $\gamma^\ast\to\pi^+\pi^-\pi^+\pi^-$ is omitted in
Ref.~\cite{lichard}. The advantage of the version exploited in our
paper is that it is chiral invariant in all sectors and hence is
justified from the point of view of basic principles. As compared
to Ref.~\cite{ecker} based on chiral amplitude, the present
analysis is alternative in that invoked are heavier resonances
characterized by two arbitrary parameters
$x_{\rho^\prime},x_{\rho^{\prime\prime}}$ instead of the higher
derivative $a_1\rho\pi$ vertex of Ref.~\cite{ecker} characterized
by three arbitrary parameters $c_{1,2,3}$. The authors of
Ref.~\cite{ecker} presented only one specific choice of three free
parameters without justifying it. They did not give the bounds for
variation of their  results against going beyond the specific
choice    made. Even with this single choice, they did not give
uncertainties nor $\chi^2/{\rm n.d.f.}$     necessary for
assessment of quality of their approach. In contrast, we clearly
state    our all assumptions and give the information necessary
for assessment of quality of    the fits. Taken literally, the
amplitude in the paper \cite{ecker} contains the direct
$\gamma^\ast\to\pi^+\pi^-$ vertex which breaks vector dominance of
the pion    form factor. In the effective chiral model used by
these authors, such breaking (which is allowed by them) can  be in
principle avoided by adjusting arbitrary parameters. The necessary
adjustment can be implemented only in zero $\rho(770)$ width
approximation.    As opposed to \cite{ecker}, we include the
demand of the vector dominance. It is important that vector
dominance can be implemented in HLS without demanding the
vanishing $\Gamma_\rho$.

One can show that HLS parameter $a$ enters the $\rho^0\to2\pi^+
2\pi^-$ decay amplitude as $a^{1/2}$. This is due to the fact that
the above amplitudes contains the factor
$$\frac{g_{\rho\pi\pi}}{f^2_\pi}=\sqrt{a}\frac{m_\rho}{2f^3_\pi},$$  provided
$g_{\rho\pi\pi}$ is expressed through $m_\rho$ and $f_\pi$
\cite{bando88}. Hence the variation around $a=2$ within 20 percent
results in variation of overall factor in cross section within the
same limits, while the difference between the measures cross
section and the calculation in HLS is clearly dynamical effect of
a stronger energy dependence than predicted in GHLS with lowest
number of derivatives. Hence, the inclusion of HLS parameter $a$
into fits will not result in any appreciable shift of the fitted
$\rho^\prime$, $\rho^{\prime\prime}$ couplings
$x_{\rho^\prime,\rho^{\prime\prime}}$.

We intentionally limit ourselves by $s^{1/2}<1$ GeV because our
goal is testing GHLS as    specific chiral model, not the study of
the $\rho$ excitations. Such rather low energy is    necessary in
order to rely on the tree  chiral amplitudes for vertices but
allowing for finite widths  for vector mesons. In fact, the
relevant invariant masses of the pion pairs are such that the
effects of the finite  widths of vector resonances  in
intermediate states are small, hence the loop effects due to
finite width are also effectively suppressed in the chosen energy
range. Hence, upon choosing above energy range we are in almost
pure situation when the tree contribution is dominant. Extending
the consideration to higher energies    in the framework of chiral
models demands inclusion of higher derivatives in     effective
chiral lagrangian and adding chiral loops. This goes far beyond
the    scope of our study. At present, the hadron physics
community  is only at the start of this very difficult road.

The inclusion of scalar resonances whose contributions may be
essential \cite{ecker,czyz08,decker94,czyz00}, deserves another
study in the chiral framework, because canonical hidden local
symmetry model is based on nonlinear realization of chiral
symmetry which does not include scalar mesons.

\section{Conclusion.}
\label{conclusion} ~

To conclude, GHLS model which includes the ground state vector and
axial vector resonances with the minimal number of derivatives
fails to explain the cross section of the reaction
$e^+e^-\to\pi^+\pi^-\pi^+\pi^-$ at energies $0.8<\sqrt{s}\leq1$
GeV. One possible way out this difficulty by including heavy
resonances $\rho^\prime$, $\rho^{\prime\prime}$ is studied here.
It is found that the contribution of these resonances is much
grater than the $\rho(770)$ contribution at $\sqrt{s}\sim1$ GeV,
and comparable with it at $\sqrt{s}\sim m_\rho$. For the sake of
simplicity, the assumption Ref.~\cite{cmd2_99} of the $a_1\pi$
dominance in the
$\rho^\prime,\rho^{\prime\prime}\to\pi^+\pi^-\pi^+\pi^-$ decays is
supposed in the present analysis. The model of similar couplings
of $\rho(770)$, $\rho^\prime$, $\rho^{\prime\prime}$ results in
qualitatively same conclusions about the fraction of
$\rho^\prime$, $\rho^{\prime\prime}$ resonances. The GHLS chiral
model used in the present work is based on the assumption of the
nonlinear realization of chiral symmetry. It would be desirable to
readdress the present issues in the framework of the chiral model
of the vector and axial vector mesons based on the linear
$\sigma$-model. This task is necessary in order to evaluate the
robustness of the figures characterizing the contributions of
heavier resonances towards various model assumptions and to reveal
the role of the intermediate states which include the widely
discussed scalar $\sigma$ meson. We hope to return to this problem
in near future.

The work is partially supported by the grants of the Russian
Foundation for Basic Research  RFBR-07-02-00093 and of the Support
of the Leading Scientific Schools NSh-5362.2006.2 and
NSh-1027.2008.2.

\appendix*
\section{The divergence of the axial vector currant in HLS model and the Adler condition.}
~

The amplitude $M^{(\gamma)}\equiv
M_{\gamma^\ast\to\pi^+_{q_1}\pi^-_{q_3}\pi^+_{q_2}\pi^-_{q_4}}$ of
direct  transition $\gamma^\ast\to\pi^+\pi^-\pi^+\pi^-$ obtained
from the lagrangian Eq.~(\ref{photon}) upon neglecting (for the
technical convenience) of the $a_1$ contribution looks as
\begin{eqnarray}
M^{(\gamma)}&=&\frac{e}{f^2_\pi}(\epsilon,q_3+q_4-q_1-q_2)+
2eg^2\times\nonumber\\&&
\left[\frac{(\epsilon,q_1-q_3)}{D_\rho(q_1+q_3)}+
\frac{(\epsilon,q_2-q_3)}{D_\rho(q_2+q_3)}
+\right.\nonumber\\&&\left.\frac{(\epsilon,q_1-q_4)}{D_\rho(q_1+q_4)}+
\frac{(\epsilon,q_2-q_4)}{D_\rho(q_2+q_4)}\right],
\label{ga4pi}\end{eqnarray} where $\epsilon$ stands for the
polarization four-vector of the virtual photon. Just  this
expression  is used in that part of Eq.~(\ref{A1}) which does not
refer to the intermediate $a_1$ meson.  The limiting expression of
the above amplitude at $m_\rho^2\to\infty$, having in mind KSRF
relation Eq.~(\ref{KSRF}), is
\begin{equation}
M^{(\gamma)}\approx\frac{e}{f^2_\pi}(\epsilon,q_1+q_2-q_3-q_4).
\label{ga4pi1}\end{equation} Notice that
\begin{equation}
(\epsilon,q_1+q_2+q_3+q_4)=0\label{transversity}\end{equation} as
the consequence of the transverse character of the (virtual)
photon. Setting, say, $q_4\to 0$ results in the expression
\begin{equation}
M^{(\gamma)}|_{q_4=0}=-\frac{2e}{f^2_\pi}(\epsilon,q_3)\not=0,\label{ga4pi2}\end{equation}
in contradiction with the Adler condition. Let us show that this
breaking of the Adler condition by the pointlike
$\gamma^\ast\to\pi^+\pi^-\pi^+\pi^-$ contribution  is the direct
consequence of the breaking of the axial current conservation by
electromagnetic field. We perform this task in the simple HLS
model neglecting the $a_1$ contribution and in the limit of
infinite $\rho$ meson mass $m_\rho\to\infty$. The inclusion of the
intermediate $\rho$ and $a_1$ resonances is straightforward
(however, technically cumbersome in case of the $a_1$
contribution) and  does not alter the above conclusion.

The hidden local symmetry  lagrangian
\cite{bando88,bando88a,bando85} looks like $ {\cal L}_{\rm
HLS}=f^2_\pi{\rm
Tr}\left(\alpha_{\bot\mu}^2+a\alpha_{\|\mu}^2\right)$, where
\begin{eqnarray}
\alpha_{\bot\mu}&=&\left(\frac{\partial_\mu\xi_R\xi^\dag_R-
\partial_\mu\xi_L\xi^\dag_L}{2i}+\right.\nonumber\\&&\left.
\frac{\xi_RR_\mu\xi^\dag_R-\xi_LL_\mu\xi^\dag_L}{2}\right),\nonumber\\
\alpha_{\|\mu}&=&\left(\frac{\partial_\mu\xi_R\xi^\dag_R+
\partial_\mu\xi_L\xi^\dag_L}{2i}+\right.\nonumber\\&&\left.
\frac{\xi_RR_\mu\xi^\dag_R+\xi_LL_\mu\xi^\dag_L}{2}-gV_\mu\right),
\label{albot}\end{eqnarray}and the kinetic energy of all vector
fields $V_\mu$, $R_\mu$, and $L_\mu$ are omitted because they are
irrelevant for the present discussion. We also assume here that
the explicit breaking of chiral symmetry necessary to make nonzero
masses of the Goldstone bosons is absent. The HLS parameter $a$ is
arbitrary, however, the convenient choice $a=2$ \cite{bando88}
results in KSRF relation Eq.~(\ref{KSRF}) and the vector meson
dominance of the pion form factor. The lagrangian  is invariant
under the transformations
\begin{eqnarray}
L_\mu&\to& g_LL_\mu g^\dag_L-i\partial_\mu g_Lg^\dag_L,\nonumber\\
R_\mu&\to& g_RR_\mu g^\dag_R-i\partial_\mu g_Rg^\dag_R,\nonumber\\
V_\mu&\to&hV_\mu h^\dag-i\partial_\mu hh^\dag,\nonumber\\
\xi_{L,R}&\to&h\xi_{L,R}g^\dag_{L,R},
\label{hlstrans}\end{eqnarray}where $g_{L,R}$ refers to the chiral
transformation, while $h$ does to the hidden gauge one. The matrix
$U$ in Eq.~(\ref{lgb}) is expressed as $U=\xi^\dag_L\xi_R$, and is
transformed according to the law $U\to g^\dag_LUg_R$. The vector
fields $V_\mu$ (corresponding to the resonances $\rho$, $\omega$,
etc) are introduced on the basis of the invariance under the truly
local hidden gauge transformation $h$, while  and external vector
fields $R_\mu$ and $L_\mu$ (corresponding to photon and weak gauge
bosons) are introduced in such a way as if ${\cal L}_{\rm HLS}$
were invariant under the  {\it local} chiral transformations
$g_{L,R}$. When the weak gauge bosons are decoupled, one has
$L_\mu=R_\mu=eQ{\cal A}_\mu$, with $Q={\rm
diag}(\frac{2}{3},-\frac{1}{3},-\frac{1}{3})$ being the charge
matrix, and ${\cal A}_\mu$ is the vector-potential of
electromagnetic field. Let us restrict ourselves  by the  sector
of non-strange mesons.  Consider now the variation of the HLS
lagrangian under space-time dependent infinitesimal chiral
transformations
$g_{L,R}(x)=1+i\epsilon^a_{L,R}(x)\frac{\tau^a}{2}$ (hereafter
$\tau^a$ stands for the standard Pauli isospin matrices) and apply
the method of Gell-Mann and Levy  to find the right and left
currents as $j^a_{\mu(R,L)}=\partial{\cal L}_{\rm
HLS}/\partial(\partial_\mu\epsilon^a_{R,L}).$ Then the  axial
vector current  $j^a_{\mu,A}=j^a_{\mu,R}-j^a_{\mu,L}$ is
\begin{eqnarray}
j^a_{\mu,A}&=&-f^2_\pi{\rm
Tr}\left[\left(\frac{\partial_\mu\xi_R\xi^\dag_R-
\partial_\mu\xi_L\xi^\dag_L}{2i}+\right.\right.\nonumber\\&&\left.\left.
e{\cal A}_\mu\frac{\xi_RQ\xi^\dag_R-\xi_LQ\xi^\dag_L}{2}\right)
\left(\xi_R\frac{\tau^a}{2}\xi^\dag_R+\right.\right.\nonumber\\&&\left.\left.
\xi_L\frac{\tau^a}{2}\xi^\dag_L\right)+a\left(\frac{\partial_\mu\xi_R\xi^\dag_R+
\partial_\mu\xi_L\xi^\dag_L}{2i}+\right.\right.\nonumber\\&&\left.\left.
e{\cal
A}_\mu\frac{\xi_RQ\xi^\dag_R+\xi_LQ\xi^\dag_L}{2}-gV_\mu\right)\times\right.\nonumber\\&&\left.
\left(\xi_R\frac{\tau^a}{2}\xi^\dag_R-
\xi_L\frac{\tau^a}{2}\xi^\dag_L\right)\right],\label{axcur}
\end{eqnarray}where $Q=\frac{\tau^3}{2}+\frac{1}{6}$. Then, taking
into account the fact that chiral  symmetry is global (constant
$\epsilon^a_{L,R}$), one finds the divergence of the right and
left currents as $\partial_\mu j^a_{\mu(R,L)}=\partial{\cal
L}_{\rm HLS}/\partial\epsilon^a_{R,L}.$ The divergence of the
axial vector found in this way is
\begin{equation}
\partial_\mu j^a_{\mu,A}=e{\cal
A}_\mu\epsilon_{3ab}j^b_{\mu,A},\label{divaxcur}\end{equation}
where $\epsilon_{abc}$ is totally antisymmetric, and
$\epsilon_{123}=1$. The divergence equation (\ref{divaxcur}) looks
like the precession (in the isotopic space) of the axial current
vector around the isovector component of the electromagnetic
field, that is, $j^3_{\mu,A}$ is conserved, while $j^\pm_{\mu,A}$
is not. Choosing the gauge
$\xi^\dag_L=\xi_R=\exp(i{\bm\pi}\cdot\frac{{\bm\tau}}{2})$, and
setting $V_\mu={\bm\rho}_\mu\cdot\frac{{\bm\tau}}{2}$ one can
obtain the soft pion expansion of the axial current to the
necessary order (up to three pions):
\begin{eqnarray}
j^a_{\mu,A}&\approx&-f_\pi\left\{\partial_\mu\pi^a-\frac{3a-4}{6f^2_\pi}
[{\bm\pi}\times[{\bm\pi}\times\partial_\mu{\bm\pi}]]^a+\right.\nonumber\\&&\left.e(a-1){\cal
A}_\mu\epsilon_{3ab}\pi^b\left(1-\frac{2{\bm\pi}^2}{3f^2_\pi}\right)-
\right.\nonumber\\&&\left.ag[{\bm\pi}\times{\bm\rho}_\mu]^a\right\}.
\label{axcur1}\end{eqnarray}In the limit of heavy $\rho$ meson its
field can be replaced by the combination
${\bm\rho}_\mu=-\frac{1}{2gf^2_\pi}[{\bm\pi}\times\partial_\mu{\bm\pi}]$,
resulting from the field equations, so that the axial current in
this limit becomes
\begin{eqnarray}
j^a_{\mu,A}&\approx&-f_\pi\left\{\partial_\mu\pi^a+\frac{2}{3f^2_\pi}
[{\bm\pi}\times[{\bm\pi}\times\partial_\mu{\bm\pi}]]^a+\right.\nonumber\\&&\left.e(a-1){\cal
A}_\mu\epsilon_{3ab}\pi^b\left(1-\frac{2{\bm\pi}^2}{3f^2_\pi}\right)\right\}.
\label{axcur2}\end{eqnarray} Taking the  matrix element of the
divergence equation (\ref{divaxcur}) relevant for
$M^{(\gamma)}|_{q_4=0}$ and  setting the HLS parameter $a=2$,  one
obtains with the help of Eq.~(\ref{axcur2})  the equation
\begin{eqnarray}
\langle\pi^+_{q_1}\pi^+_{q_2}\pi^-_{q_3}|\partial_\mu
j^-_{\mu,A}|\gamma^\ast\rangle&=&\frac{2e}{3f_\pi}(\epsilon,2q_3-q_1-q_2)=\nonumber\\&&
\frac{2e}{f_\pi}(\epsilon,q_3).\label{divaxcur2}\end{eqnarray}
[One should  kept in mind Eq.~(\ref{transversity}) taken at
$q_4=0$.] Allowing for the fact that to the leading order the
axial current and the gradient of the pion field are related by
the factor $-f_\pi$ [see Eq.~(\ref{axcur2})], one can see that the
breaking of the Adler condition expressed by Eq.~(\ref{ga4pi2}) is
the direct consequence of non-conservation of the axial current by
external electromagnetic field expressed by Eq.~(\ref{divaxcur2}).
The cases of vanishing of other three pion momenta $q_1$, $q_2$,
or $q_3$ are treated in the same manner, by taking the relevant
matrix elements of the axial current divergence equation.



\begin{thebibliography}{99}
\bibitem{fn1}
In fact, the use of  the expression for the four pion production
amplitude which includes the $\rho(770)-\rho(1450)-\rho(1700)$
mixing proposed in Ref.~\cite{ach97} results in no qualitative
difference in the role of heavy resonance at $\sqrt{s}\leq1$ GeV
as compared to the fixed width, no mixing form adopted in the
present paper.
\end{thebibliography}
\end{document}